\documentclass[%
 aip,
 amsmath,amssymb,
 reprint,%
]{revtex4-1}
\usepackage{amsfonts}
\usepackage{amsmath}
\usepackage{amssymb}
\usepackage{graphicx}
\usepackage{epstopdf}
\usepackage{mathrsfs}
\usepackage{color}
\usepackage{bbold}
\usepackage{CJK}
\usepackage{times}
\usepackage[colorlinks, citecolor=red]{hyperref}

\begin{document}

\title{Very large thermal rectification in ferromagnetic insulator-based superconducting tunnel junctions}

\author{F. Giazotto}
\email{francesco.giazotto@sns.it}
\affiliation{NEST Istituto Nanoscienze-CNR and Scuola Normale Superiore, I-56127 Pisa, Italy}

\author{F. S. Bergeret}
\email{fs.bergeret@csic.es}
\affiliation{Centro de Fisica de Materiales (CFM-MPC), Centro Mixto CSIC-UPV/EHU, Manuel de
Lardizabal 5, 20018  Donostia-San Sebastian, Spain}
\affiliation{Donostia International Physics Center (DIPC), Manuel de
Lardizabal, 4, 20018, Donostia San Sebastian, Spain}

\begin{abstract}
We investigate electronic thermal rectification in ferromagnetic insulator-based superconducting tunnel junctions. Ferromagnetic insulators coupled to superconductors are known to induce sizable spin splitting in the superconducting density of states,  and also  lead to  efficient spin filtering if operated as tunnel barriers. 
The combination of spin splitting and spin filtering is shown to yield a substantial \emph{self-amplification} of the electronic heat diode effect due to breaking of the electron-hole symmetry in the system which is added  to the thermal  asymmetry of the junction. Large spin splitting and large spin polarization  can potentially lead to thermal rectification efficiency exceeding $\sim 5\times 10^4 \%$ for realistic parameters  in a suitable temperature range, thereby outperforming up to a factor of $\sim 250$ the heat diode effect achievable with conventional superconducting tunnel junctions. These results could be relevant for improved mastering of the heat currents in innovative phase-coherent caloritronic nanodevices, and for enhanced thermal management of quantum circuits at the nanoscale.  
\end{abstract}

\maketitle
A \emph{thermal} rectifier, or heat diode, \cite{roberts2011review,li2004thermal} is a device in which the heat current depends on the sign  and the amplitude of the temperature gradient imposed across it.  
The implementation of efficient heat diodes
would represent a breakthrough in the realization of improved  thermal circuits  for cooling \cite{giazotto2006opportunities}, thermal isolation,  energy harvesting, radiation sensing, and several other applications \cite{fornieri2017towards}. 
Recently, the control of   thermal transport  at the nanoscale has been attracting great interest  \cite{fornieri2017towards,martinez2014coherent,giazotto2006opportunities,fornieri2015electronic}.
From the theoretical side, strong effort has been put to conceive thermal rectification setups dealing with phonons \cite{wu2009sufficient,segal2008single,li2006negative,terraneo2002controlling}, electrons \cite{lopez2013nonlinear,ren2013anomalous,bours2019phase,ruokola2011single,ruokola2009thermal,kuo2010thermoelectric,fornieri2014normal,martinez2013efficient,giazotto2013thermal,fornieri2015electronic,goury2019reversible} and photons \cite{ben2013phase}.
Experimentally, promising results  were obtained so far in the context of electronic \cite{martinez2015rectification,senior2020heat,scheibner2008m} and phononic \cite{chang2006solid,kobayashi2009oxide,tian2012novel} heat transport.
Rectification of electronic heat currents has been studied  in several types of  tunneling junctions between different materials such as, for instance,  normal metals \cite{fornieri2014normal}, Josephson junctions \cite{martinez2013efficient}, and superconductor-normal metal structures\cite{giazotto2013thermal,fornieri2015electronic}. 
In all cases,   heat rectification  stems from thermal asymmetry of the structure in the direction of the current flow.  
In  conventional superconducting  tunnel junctions this asymmetry together with the highly non-linear temperature dependence of the density of states 
leads to a heat rectification up to $\sim 800\%$ in Josephson junctions \cite{martinez2013efficient}.  

In this letter we study thermal  rectification effects in systems based on    ferromagnetic insulator (FI)  superconducting tunnel junctions. 
The combination of spin splitting and spin polarization induced by FIs   
breaks the electron-hole symmetry  of the electronic transport. This electron-hole symmetry breaking (EHSB) together with non-linear temperature dependence of the superconducting spectral properties  leads to an unrivalled heat diode effect with a suitable choice of realistic  parameters.  Specifically, a rectification efficiency exceeding $\sim 5\times 10^4 \%$ can be obtained, outperforming up to a factor of $\sim 250$  that achievable in conventional superconducting tunnel junctions. These results could be relevant for improved thermal isolation of cryogenic quantum circuitry at the nanoscale.

\begin{figure}[t!]
\includegraphics[width=0.9\columnwidth]{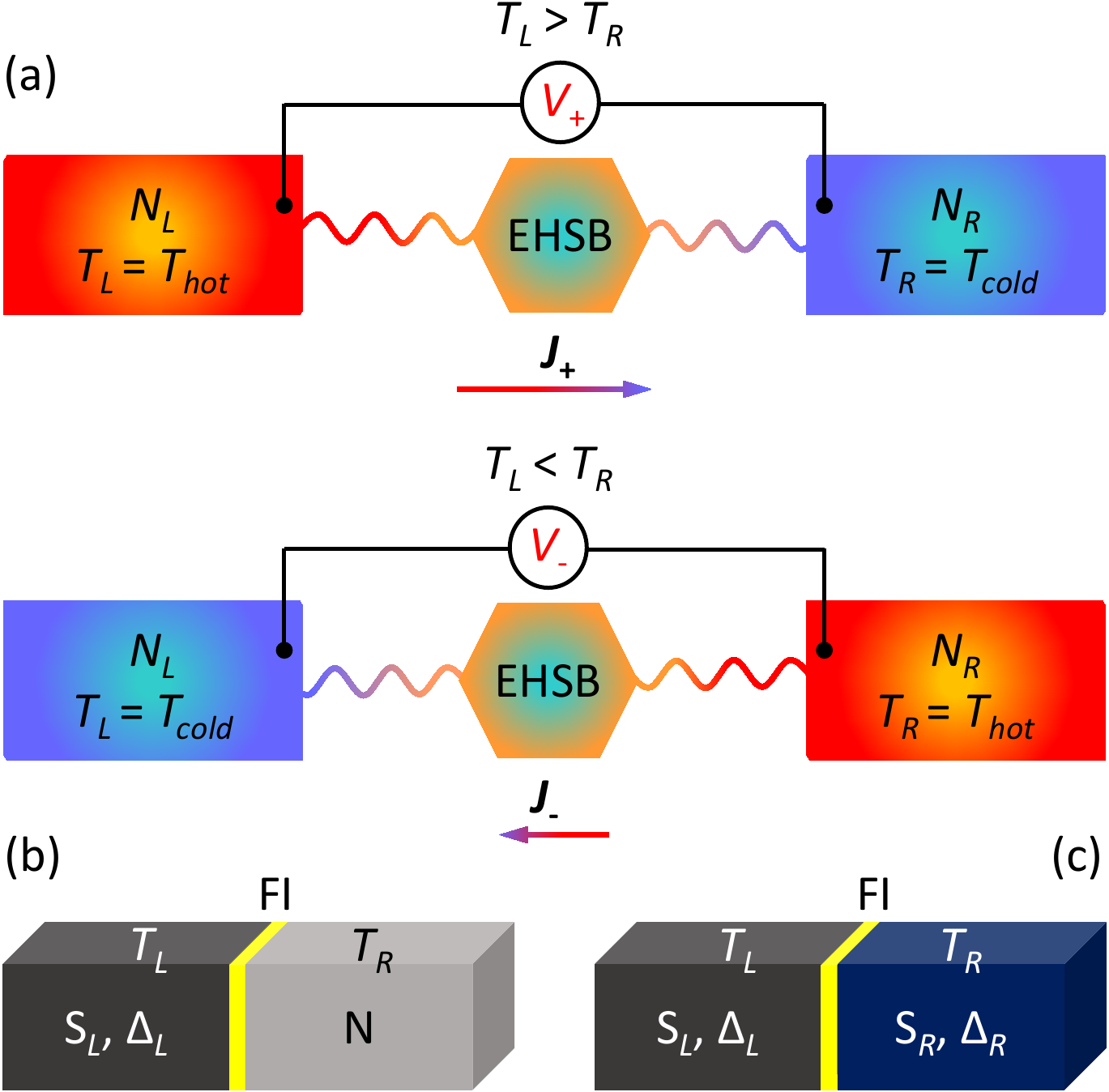}
\caption{(a) Sketch of two electronic reservoirs (either normal or superconducting) with different temperature-dependent density of states $N_{L,R}$ residing at temperature $T_{L,R}$. The reservoirs are coupled  to an element that breaks electron-hole symmetry.  This  leads to a thermoelectric response, and enhanced  \emph{thermal} rectification.
$J_+$ and $J_-$ represent the heat flow in the
forward ($T_L>T_R$) and reverse ($T_L<T_R$) thermal-bias configuration, respectively, while $V_+$ and $V_-$ denote the corresponding thermovoltages developed across the system.
(b) Prototypical ferromagnetic insulator (FI)-based superconductor (S$_L$)-normal metal (N) and (c) superconductor (S$_L$)-superconductor (S$_R$) tunnel junctions. The FI is inserted in the structures as a tunnel barrier so to induce both a spin splitting in the density of states of the superconductor, and to provide a spin filtering effect.
$\Delta_L$($\Delta_R)$ denotes the energy gap in S$_L$(S$_R$).
}
\label{setup}
\end{figure}

\emph{Setup and basic equations} Figure \ref{setup}(a) schematizes the generic system under investigation which consists of two electronic reservoirs (either superconducting or normal metallic) residing at temperature $T_{L,R}$.  The  density of states $N_{L,R}$ depends  on the temperature and applied exchange field. 
The two electrodes  are  electrically coupled  via an  EHSB mechanism.  The latter gives rise to a thermoelectric response in the system, and to a thermovoltage amplitude $V_+$($V_-$) developed for $T_L>T_R$($T_L<T_R$). The heat current flowing in the forward thermal-bias configuration is denoted with $J_+$ whereas $J_-$ denotes the one flowing in the reverse thermal-bias configuration. As we shall show, the presence of EHSB in the system yields a \emph{self-amplification} of thermal rectification efficiency up to unparalleled values for suitable parameters of the structure, and proper thermal bias conditions.  The EHSB mechanism can be achieved  by placing a S-FI building block in two possible configurations: a S$_L$FIN , [Fig. \ref{setup}(b)] and a S$_L$FIS$_R$ , [Fig. \ref{setup}(c)] tunnel junctions.  The presence of the FI layer yields both spin splitting of the density of states in $S_L$  and spin filtering at the barrier. The combination results in   the EHSB mechanism \cite{ bergeret2018colloquium,heikkila2019thermal,ozaeta2014predicted,machon2013nonlocal,machon2014giant,kolenda2016observation,linder2015superconducting}.

The interaction between the spin of conducting electrons in the superconductor and the localized magnetic moments in
the adjacent  FI leads to  an effective exchange interaction ($h_{exc}$) in the superconductor. This field  decays away from the S/FI interface  over the superconducting coherence length $\xi_0$ \cite{tokuyasu1988proximity}.
Yet, we assume that the superconducting layer thickness is  smaller than $\xi_0$, so that the induced $h_{exc}$  in the superconductor by FI is spatially homogeneous.
In this situation, the spin-dependent normalized density of states of the superconductor is simply given by 
$N^{\uparrow, \downarrow} (E)=\frac{1}{2}|{\rm Re}[\frac {E+i\Gamma \pm h_{\rm exc}}{\sqrt{\left(E+i\Gamma\pm h_{\rm exc}\right)^2-\Delta^2}}]|$,
where $\Gamma$ is the Dynes parameter, and $\Delta$ is the superconducting gap which depends on $T$ and $h_{\rm exc}$ via the  self-consistency equation
$\ln\left(\frac{\Delta_0}{\Delta}\right)=\int_0^{\hbar\omega_D}dE\frac{f_+(E)+f_-(E)}{\sqrt{E^2+\Delta^2}}$,
where $f_{\pm}(E)=\left\{1+\textrm{exp}[\frac{1}{k_B T}(\sqrt{E^2+\Delta^2}\mp h_{exc})]\right\}^{-1}$, $\omega _D$ is the Debye frequency of the superconductor, $\Delta_0=1.764k_BT_c$ is the zero-temperature, zero-exchange field superconducting pairing potential, $T_c$ is the critical temperature, and $k_B$ is the Boltzmann constant. The parameter $\Gamma$ accounts for the broadening of the coherent peaks in the density of states due to inelastic scattering, and for an ideal superconductor $\Gamma\rightarrow 0^+$ \cite{dynes1984tunneling}. In all the calculations we set $\Gamma=10^{-4}\Delta_0$, unless differently stated.

We are interested in both  the DC charge and electronic heat  currents  through the  junctions which  are given  by \cite{ozaeta2014predicted} 
$I=\frac{1}{eR_T}\int_{-\infty}^{\infty} dE\left[N_++PN_-\right][f_L(V,T_L)-f_R(T_R)]$, and $J=\frac{1}{eR_T}\int_{-\infty}^{\infty} dE(E+eV)\left[N_++PN_-\right][f_L(V,T_L)-f_R(T_R)]$, respectively. 
Here, $R_t$ is the normal-state tunneling resistance of the  junction,  $N_\pm=(N_L^\uparrow N_R^\uparrow\pm N_L^\downarrow N_R^\downarrow)$, and $0\leq P\leq 1$ is the barrier spin polarization provided by the FI layer \cite{moodera2007phenomena}.
Moreover,  $f_{L}(V,T_L)=[1+\textrm{exp}[(E+eV)/k_BT_L]]^{-1}$ and $f_{R}(T_R)=[1+\textrm{exp}(E/k_BT_R)]^{-1}$ are the equilibrium quasiparticle distribution functions, and $e$ is the electron charge.  
In principle the expressions for both currents may contain a  phase-dependent term 
 \cite{giazotto2012josephson,martinez2014coherent,fornieri2017towards}. However,   since  our system exhibits a  thermovoltage  across the junction the phase  becomes time-dependent and  thereby not contributing to DC transport. 
It is also worth emphasizing that the values of both spin polarization and spin splitting can be extracted from  experiments, as demonstrated  in several works \cite{meservey1994spin,hao1990spin,li2013observation,li2013superconducting,xiong2011spin,liu2013electrostatic,wolf2014spin,strambini2017revealing,de2018toward,rouco2019charge}.

We assume that both electronic reservoirs are thermalized with  well defined temperatures $T_{L,R}$.  
 In the \emph{forward} thermal bias configuration [see Fig.
\ref{setup}(a)], a thermal gradient is intentionally created at
the junction by setting $T_L$ = $T_{hot}>T_R$ = $T_{cold}$, which leads to
a total heat flux $J_+$ through the system. 
In the \emph{reverse} thermal bias configuration, the heat gradient
is inverted so that $T_L = T_{cold} < T_R = T_{hot}$  yielding a total
heat current $J_-$ flowing from the right to the left electrode.
 The electronic thermal rectification efficiency is defined as 
 $R(\%)=100\times (|J_+|-|J_-|)/|J_-|$,
such that  the absence of heat rectification corresponds to  $R=0$  and  $R>0$ indicates a preferential heat flow from the left  to the right side of the junction.
In order to determine the  thermal rectification, for example in the forward thermal bias configuration,  for arbitrary  $T_L > T_R $   we need first  to determine the thermovoltage $V_+$ across the junction by solving the equation $I(V_+,T_L,T_R)=0$.  Then, the obtained thermovoltage is used to compute the corresponding heat current $J_+(V_+,T_L,T_R)$ flowing through the junction. In the reverse thermal bias condition, $T_R> T_L $   , the same procedure is performed  to determine $V_-$ and $J_-$. Following this procedure we determine $R$  in both junction setups depicted in Figs.
\ref{setup}(b-c).

\begin{figure}[t!]
\includegraphics[width=1\columnwidth]{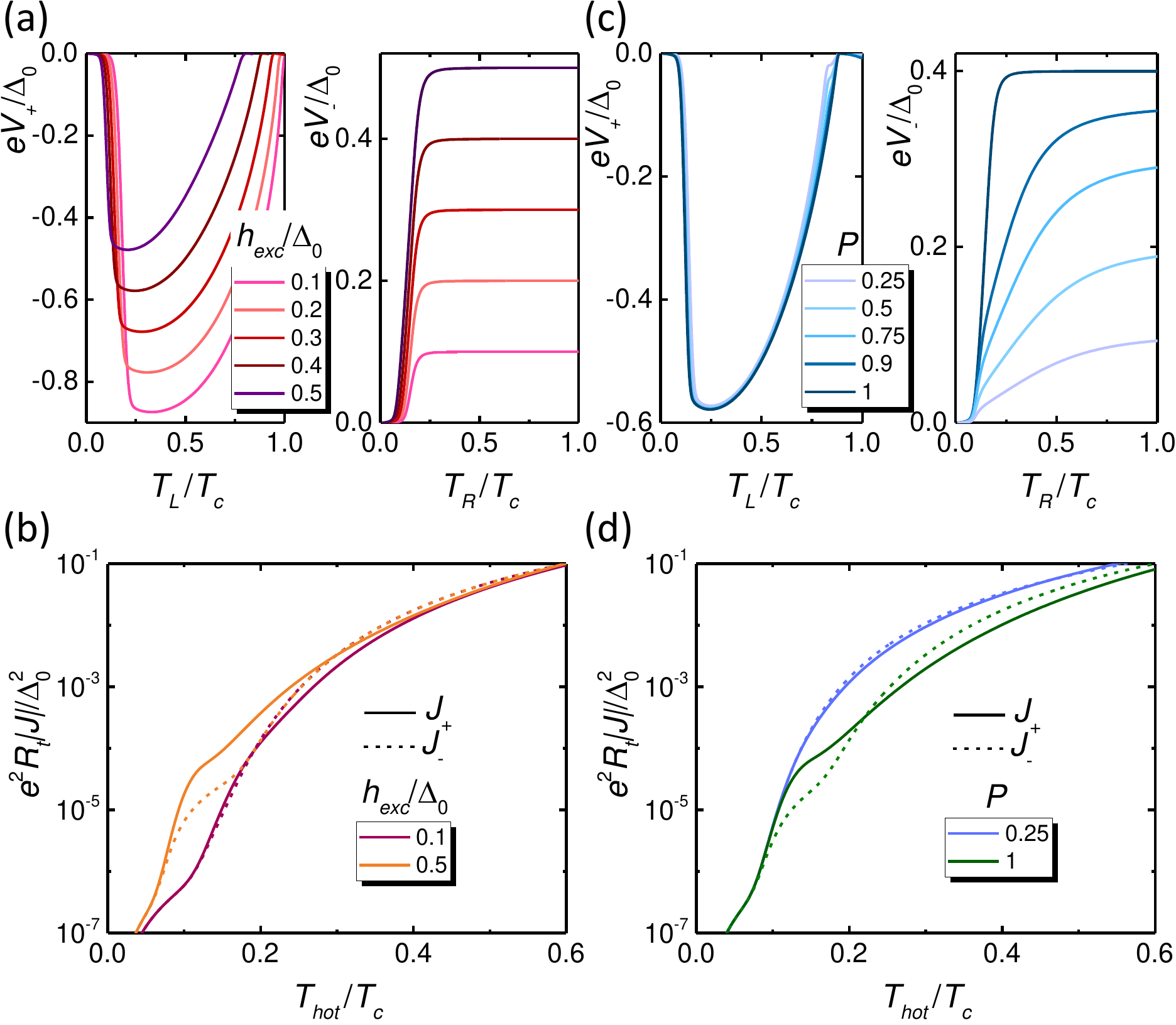}
\caption{(a) Thermovoltage $V_+$ vs $T_L$ (left panel) and  $V_-$  vs $T_R$ (right panel) calculated for a S$_L$FIN junction at $P=1$ and $T_R=0.01T_c$ (left panel) and $T_L=0.01T_c$ (right panel)  for different values of the exchange field $h_{exc}$.
(b) Absolute value of the heat current $|J|$ vs $T_{hot}$ flowing through a S$_L$FIN junction calculated at $P=1$ and $T_{cold}=0.01T_c$ for two selected values of $h_{exc}$.  
(c) $V_+$ vs $T_L$ (left panel) and  $V_-$  vs $T_R$ (right panel) calculated for a S$_L$FIN junction at $h_{exc}=0.4\Delta_0$ and $T_R=0.01T_c$ (left panel) and $T_L=0.01T_c$ (right panel)  for different values of barrier polarization $P$.
(d) Absolute value of the heat current $|J|$ vs $T_{hot}$ flowing through a S$_L$FIN junction calculated at $h_{exc}=0.4\Delta_0$ and $T_{cold}=0.01T_c$ for two selected values of $P$.  
}
\label{thermovoltage}
\end{figure}
\emph{S$_L$FIN tunnel junction} It is instructive to start our discussion by analyzing the setup of Fig. \ref{setup}(b) , the  S$_L$FIN superconducting tunnel junction. A thermal bias across the structure leads to a thermovoltage $V_{\pm}$ which depends on the sign of the thermal gradient itself, and stems from EHSB in the junction \cite{giazotto2015ferromagnetic} [see Fig. \ref{setup}(a)]. 
In particular, the thermovoltage $V_+$ vs $T_L$ (at $T_R=0.01T_c$) is shown in the left panel of Fig. \ref{thermovoltage}(a) whereas $V_-$  vs $T_R$ (at $T_L=0.01T_c$) is shown in the right panel of the same figure, both evaluated at $P=1$ for selected values of the exchange field $h_{exc}$. 
 Beside the substantial difference between the thermovoltage amplitudes $V_+$ and $V_-$ for the same exchange field, $V_+$ is a non-monotonic function of $T_L$, vanishing when the superconducting pairing potential goes to zero,  while $V_-$ monotonically increases with $T_R$, saturating  at  the asymptotic value $eV_-=h_{exc}$ at large temperature \cite{giazotto2015ferromagnetic}.  This sizable difference between the  thermovoltages has a direct  impact  on  the corresponding heat currents $J_{\pm}$ flowing through the junction. These are  shown in Fig. \ref{thermovoltage}(b) for two given  values of the exchange field.  For the larger value of  $h_{exc}$ the difference between the forward and reverse  heat currents is increased  thereby leading to enhanced thermal rectification in the S$_L$FIN junction. This difference appears to be particularly pronounced for $0.1T_c\lesssim T_{hot}\lesssim 0.2T_c$, which is indeed the temperature range where the maximum thermovoltage amplitudes develop across the junction [{\it cf.} Fig. \ref{thermovoltage}(a)].

Left panel of Fig. \ref{thermovoltage}(c) displays $V_+$ vs $T_L$ (at $T_R=0.01T_c$) while the right one shows $V_-$  vs $T_R$ (at $T_L=0.01T_c$) both calculated at $h_{exc}=0.4\Delta_0$ for a different values of  $P$. The thermovoltage $V_+$ is negligibly affected by the polarization showing a shape very similar to that obtained for the same value of $h_{exc}$ in panel (a). By contrast, $V_-$ is strongly affected by $P$, and becomes larger by increasing the barrier polarization. Also in the present case of finite barrier polarization the large difference of thermovoltage amplitudes deeply affect the corresponding heat currents flowing through the structure, as shown in Fig. \ref{thermovoltage}(d) for two selected values of $P$. A  large   barrier polarization ($P\sim1$) strongly enhances the difference between forward and reverse heat currents, therefore leading to a sizable heat diode effect in the junction.

\begin{figure}[t!]
\includegraphics[width=1\columnwidth]{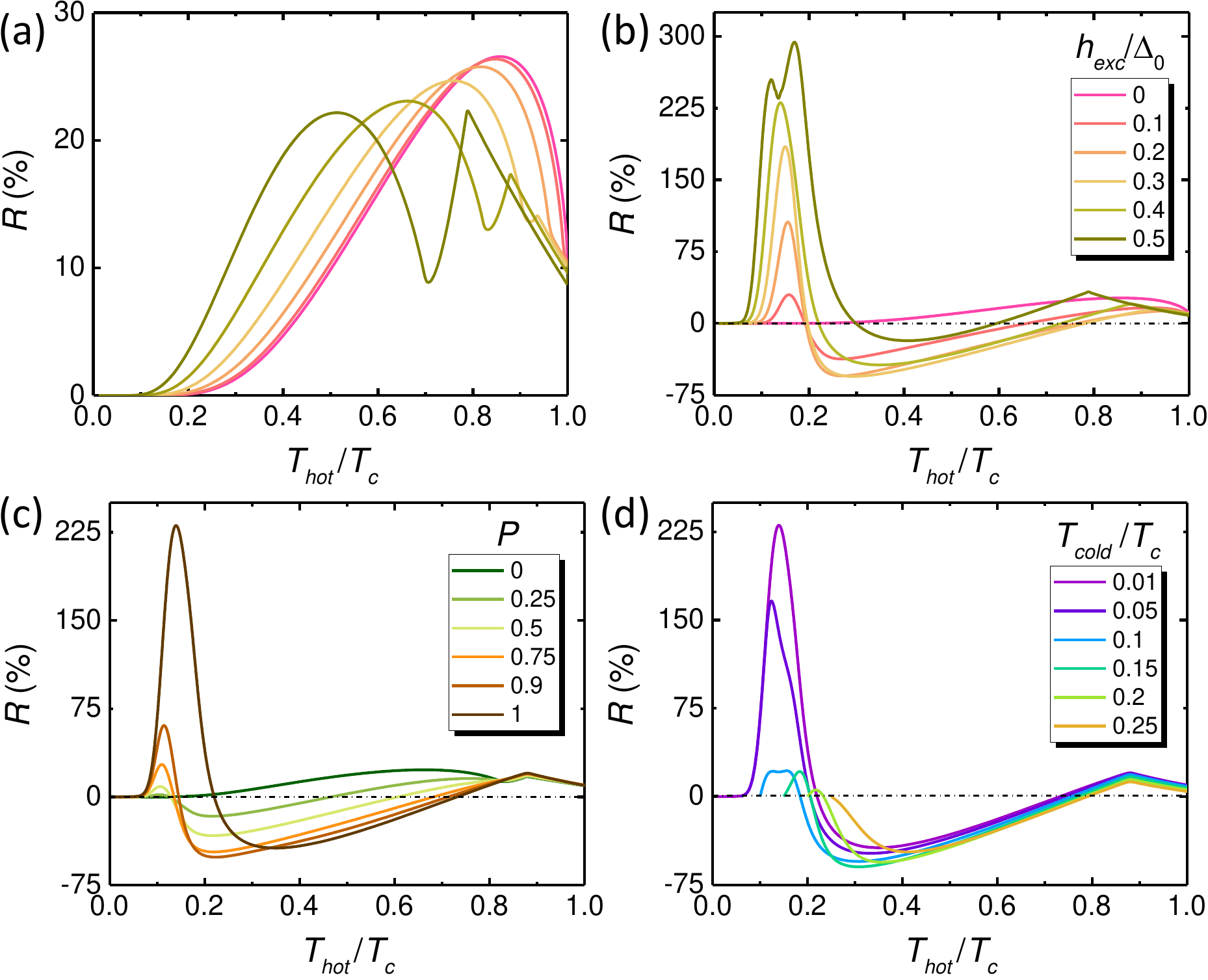}
\caption{Thermal rectification coefficient  $R$ for the S$_L$FIN junction.  (a) Thermal rectification coefficient $R$ vs $T_{hot}$ calculated at $T_{cold}=0.01T_c$ and $P=0$ for different values of the exchange field $h_{exc}$. (b) The same as in panel (a) but for barrier polarization $P=1$.
(c) $R$ vs $T_{hot}$ calculated at $T_{cold}=0.01T_c$ and $h_{exc}=0.4\Delta_0$ for a few values of $P$.
(d) $R$ vs $T_{hot}$ calculated at $h_{exc}=0.4\Delta_0$ and $P=1$ for several values of $T_{cold}$. Dashed lines indicate $R=0$.
}
\label{SFIN}
\end{figure}

In Figure \ref{SFIN} we show the  rectification efficiency for the $S_LFIN$ junction. Specifically,  \ref{SFIN}(a) shows the heat rectification efficiency $R$ vs $T_{hot}$ calculated for zero barrier polarization at $T_{cold}=0.01T_c$ for different  values of $h_{exc}$. The absence of  spin polarization at the barrier leads to a zero  thermoelectric voltage.   The increase of $h_{exc}$ yields a slight reduction of $R$ compared to that of a conventional SIN tunnel junction (i.e., for $h_{exc}=0$) \cite{giazotto2013thermal,martinez2013efficient,fornieri2015electronic}, allowing to obtain a maximum rectification efficiency of $\sim 22\%$ for $h_{exc}=0.5\Delta_0$ at $T_{hot}\sim 0.5 T_c$. 
The above heat rectification reduction stems from a larger contribution to the  heat transport for electrons  with energy  close to the Fermi level  at large $h_{exc}$. This contribution restore  the thermal symmetry of the junction, and hence reduces the heat rectification.  In addition,  there are additional features appearing in the rectification characteristics at higher temperatures which occur when  superconductivity is quenched due to the presence of a finite exchange field. In short, Fig. \ref{SFIN}(a) demonstrates that spin-splitting alone cannot improve thermal rectification with respect to conventional superconducting junctions.

However, if the barrier spin-polarization is finite the situation changes drastically.  The role of finite $h_{exc}$ at $P=1$ is shown in Fig. \ref{SFIN}(b). In particular, the thermoelectric response of the junction is increased in the presence of high exchange field allowing to obtain both large heat rectification in the forward thermal-bias configuration for $h_{exc}=0.5\Delta_0$ (i.e., up to $\sim 290\%$) and a sizable  contribution in the reverse thermal-bias configuration (around $\sim -55\%$ for $h_{exc}=0.3\Delta_0$) . 
All the above results prove that the presence of an EHSB mechanisms in the S$_L$FIN junction leads to a substantial \emph{self-amplification} of the heat diode efficiency which now obtains values which are larger by more than a factor of $\sim 10$ than those typically achievable in conventional superconducting tunnel junctions \cite{giazotto2013thermal,martinez2013efficient,fornieri2015electronic}. 

The impact of a finite barrier polarization at $h_{exc}=0.4\Delta_0$  is displayed in Fig. \ref{SFIN}(c) for different values of $P$. The effect is similar to that caused by an increasing exchange field [see Fig. \ref{SFIN}(c)], and shows the relevance of a large spin polarization in order to achieve a sizable heat rectification. For instance, $R$ turns out to be suppressed by more than a factor of $\sim 20$ if $P$ is reduced down to $50\%$.
Finally, the effect of the smaller temperature $T_{cold}$ onto $R$ is displayed in Fig. \ref{SFIN}(d) as a function of $T_{hot}$. We notice, in particular, the strong suppression of $R$ occurring by increasing $T_{cold}$: for instance, $R$ is suppressed by roughly one order of magnitude at $T_{cold}=0.1T_c$. This  emphasizes the requirement of a sufficiently low $T_{cold}$ in order to achieve large  rectification effects.

\begin{figure}[t!]
\includegraphics[width=1\columnwidth]{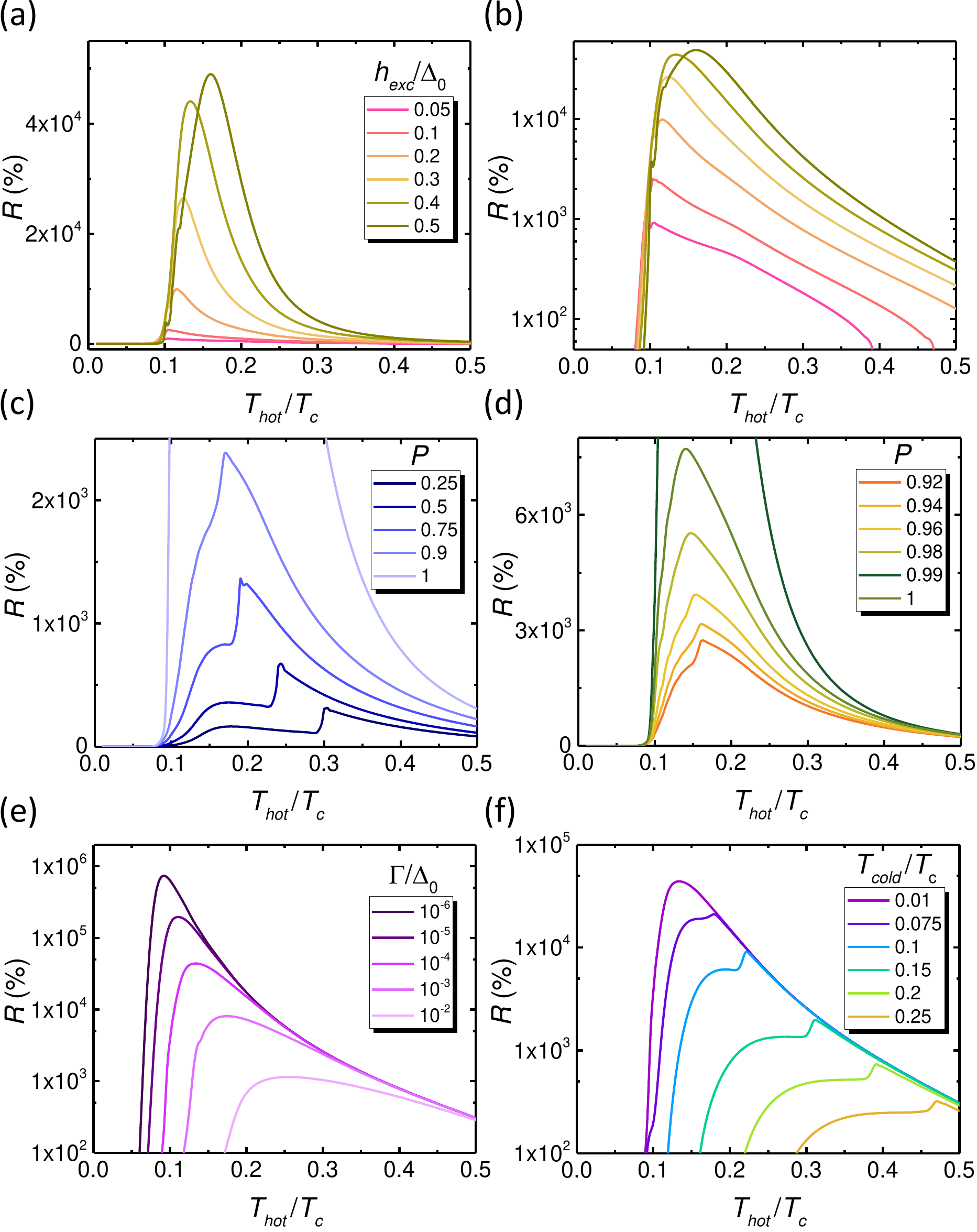}
\caption{ Thermal rectification coefficient  $R$ for the S$_L$FIS$_R$ junction.  (a)  $R$ vs $T_{hot}$ calculated at $T_{cold}=0.01T_c$ and $P=1$ for different values of the exchange field $h_{exc}$. (b) The same as in panel (a) but shown on a logarithmic scale.
(c) $R$ vs $T_{hot}$ calculated at $T_{cold}=0.01T_c$ and $h_{exc}=0.4\Delta_0$ for a few values of $P$.
(d) $R$ vs $T_{hot}$ calculated at  $T_{cold}=0.01T_c$ and $h_{exc}=0.4\Delta_0$ for several high values of barrier polarization in the range $0.92\leq P\leq 1$. 
(e) $R$ vs $T_{hot}$ calculated at $T_{cold}=0.01T_c$, $h_{exc}=0.4\Delta_0$, and $P=1$ for a few values of $\Gamma$.
(f) $R$ vs $T_{hot}$ calculated at $h_{exc}=0.4\Delta_0$ and $P=1$ for different values of $T_{cold}$.}
\label{SFIS}
\end{figure}

\emph{S$_L$FIS$_R$ tunnel junction} We  now discuss the heat diode effect in the junction setup sketched in Fig. \ref{setup}(c). For simplicity we assume the two superconductors are identical such that they have  the same zero-temperature, zero-exchange field energy gap $\Delta_0$.  To maximize the thermal asymmetry,  we also assume  that only the density of states of the left electrode ($N_L$) is affected by $h_{exc}$ so to break the thermal symmetry of the system. Such an asymmetry can be achieved by inserting a very thin non-magnetic oxide layer at the $FI/S_R$ interface\cite{hao1990spin}.  This implies that $N_R^\uparrow=N_R^\downarrow$.  We now define the rectification efficiency as $R=100\times (|J_-|-|J_+|)/|J_+|$ so to easily compare the two different junction setups since, as we shall show, $J_-$ is typically much larger than $J_+$ in the S$_L$FIS$_R$ junction.
Figures \ref{SFIS}(a) and (b) show the impact of a finite $h_{exc}$ on $R$ at $P=1$. In particular, we note the very large thermal rectification which can be obtained, up to $\sim 5\times 10^4\%$ for $h_{exc}=0.5\Delta_0$ at $T_{hot}\simeq 0.15 T_c$. This  value is  $\sim 250$ times larger than the one achievable in conventional all-superconducting tunnel junctions made with superconductors with different energy gaps \cite{martinez2013efficient,fornieri2015electronic}. Yet, even for moderate $h_{exc}$ values (i.e., $h_{exc}=0.2\Delta_0$) $R$ can reach  values as large as $10^4\%$ in the suitable $T_{hot}$ range. The above results demonstrate  the effectiveness of S$_L$FIS$_R$ tunnel junctions to achieve very high thermal rectification efficiency.

The role of $P$ is displayed in Figs. \ref{SFIS}(c) and (d), and reveal the increased robustness of the S$_L$FIS$_R$ setup with respect to the  S$_L$FIN one in terms of limited barrier polarization. For example, for a moderate spin polarization, $P=75\%$, we still get a sizable $R\simeq 1370\%$ which is about two times larger than the maximum value achievable in conventional SIS tunnel junctions \cite{martinez2013efficient,fornieri2015electronic}. 
Nowadays, for instance, state-of-the-art ferromagnetic europium (Eu) chalcogenides tunnel barriers can provide spin polarization close to  $100\%$\cite{moodera2010frontiers} which would make thermal rectification efficiencies larger than $\sim 7.5\times10^3\%$ readily available in superconducting tunnel junction setups operating at cryogenic temperatures.

The impact of non-idealities of the junction is shown in Fig. \ref{SFIS}(e) where $R$ is plotted against $T_{hot}$ for a few selected values of $\Gamma$. From a quantitative point of view,
 thermal rectification efficiency  turns out to be less effective the larger the value of $\Gamma$. In particular, for a sizable $\Gamma=10^{-2}\Delta_0$, the maximum of $R$ is reduced down to $\sim 1140\%$ at $T_{hot}\simeq 0.25T_c$. This fact emphasizes the requirement of high-quality tunnel junctions in order to preserve a substantial heat diode effect. Finally, Fig. \ref{SFIS}(f) shows  how $T_{cold}$ affects the rectification efficiency. Analogously to the S$_L$FIN setup, by increasing $T_{cold}$ deeply suppresses the $R$ coefficient, although in a reduced way. In particular, for $T_{cold}=0.1T_c$ the rectification efficiency turns out to be decreased by almost a factor of $5$ with respect to the lowest temperature, suggesting that the S$_L$FIS$_R$ junction is more efficient as a heat diode at higher temperatures than the S$_L$FIN setup.
 
 In summary, we have  demonstrated theoretically the occurrence of unparalleled thermal diode effect  in superconducting tunnel junctions with  ferromagnetic insulators. In particular, thermal rectification efficiency above $\sim 5\times 10^4\%$ could be achieved for realistic material parameters in a suitable temperature range. Such a heat rectifier efficiency exceeds by a factor of $\sim 250$ the one obtained  with conventional superconducting tunnel junctions.  
 Ideal materials for the heat rectifier  are  europium chalcogenides layers (EuO and EuS), for which values of $P$ ranging from $80\%$ up to $\sim 100\%$ have been
reported \cite{moodera1988electron,hao1990spin,santos2004observation,moodera2007phenomena,moodera1993variation,santos2008determining,miao2009controlling,li2013superconducting}, in combination with Al superconducting thin films \cite{strambini2017revealing,de2018toward,rouco2019charge}. Yet, 
very large spin-filtering has been reported in GdN barriers as well, \cite{senapati2011spin,pal2013electric,pal2014pure} with polarizations as large as $97\%$ at low temperature.
Our finding might be relevant for enhanced control of heat current in phase-coherent caloritronic devices \cite{fornieri2017towards,martinez2014coherent} as well as for general thermal management of nanoscale quantum circuits \cite{giazotto2006opportunities}.

The authors acknowledge the European Union’s Horizon 2020 research and innovation programme under the grant agreement No.  800923-SUPERTED and the  Spanish Ministerio de Ciencia e Innovacion (MICINN) through the   Project  FIS2017-82804-P, for partial financial support. 


\begin{thebibliography}{58}%
\makeatletter
\providecommand \@ifxundefined [1]{%
 \@ifx{#1\undefined}
}%
\providecommand \@ifnum [1]{%
 \ifnum #1\expandafter \@firstoftwo
 \else \expandafter \@secondoftwo
 \fi
}%
\providecommand \@ifx [1]{%
 \ifx #1\expandafter \@firstoftwo
 \else \expandafter \@secondoftwo
 \fi
}%
\providecommand \natexlab [1]{#1}%
\providecommand \enquote  [1]{``#1''}%
\providecommand \bibnamefont  [1]{#1}%
\providecommand \bibfnamefont [1]{#1}%
\providecommand \citenamefont [1]{#1}%
\providecommand \href@noop [0]{\@secondoftwo}%
\providecommand \href [0]{\begingroup \@sanitize@url \@href}%
\providecommand \@href[1]{\@@startlink{#1}\@@href}%
\providecommand \@@href[1]{\endgroup#1\@@endlink}%
\providecommand \@sanitize@url [0]{\catcode `\\12\catcode `\$12\catcode
  `\&12\catcode `\#12\catcode `\^12\catcode `\_12\catcode `\%12\relax}%
\providecommand \@@startlink[1]{}%
\providecommand \@@endlink[0]{}%
\providecommand \url  [0]{\begingroup\@sanitize@url \@url }%
\providecommand \@url [1]{\endgroup\@href {#1}{\urlprefix }}%
\providecommand \urlprefix  [0]{URL }%
\providecommand \Eprint [0]{\href }%
\providecommand \doibase [0]{http://dx.doi.org/}%
\providecommand \selectlanguage [0]{\@gobble}%
\providecommand \bibinfo  [0]{\@secondoftwo}%
\providecommand \bibfield  [0]{\@secondoftwo}%
\providecommand \translation [1]{[#1]}%
\providecommand \BibitemOpen [0]{}%
\providecommand \bibitemStop [0]{}%
\providecommand \bibitemNoStop [0]{.\EOS\space}%
\providecommand \EOS [0]{\spacefactor3000\relax}%
\providecommand \BibitemShut  [1]{\csname bibitem#1\endcsname}%
\let\auto@bib@innerbib\@empty
\bibitem [{\citenamefont {Roberts}\ and\ \citenamefont
  {Walker}(2011)}]{roberts2011review}%
  \BibitemOpen
  \bibfield  {author} {\bibinfo {author} {\bibfnamefont {N.~A.}\ \bibnamefont
  {Roberts}}\ and\ \bibinfo {author} {\bibfnamefont {D.}~\bibnamefont
  {Walker}},\ }\href@noop {} {\bibfield  {journal} {\bibinfo  {journal} {Int.
  J. Therm. Sci.}\ }\textbf {\bibinfo {volume} {50}},\ \bibinfo {pages} {648}
  (\bibinfo {year} {2011})}\BibitemShut {NoStop}%
\bibitem [{\citenamefont {Li}, \citenamefont {Wang},\ and\ \citenamefont
  {Casati}(2004)}]{li2004thermal}%
  \BibitemOpen
  \bibfield  {author} {\bibinfo {author} {\bibfnamefont {B.}~\bibnamefont
  {Li}}, \bibinfo {author} {\bibfnamefont {L.}~\bibnamefont {Wang}}, \ and\
  \bibinfo {author} {\bibfnamefont {G.}~\bibnamefont {Casati}},\ }\href@noop {}
  {\bibfield  {journal} {\bibinfo  {journal} {Phys. Rev. Lett.}\ }\textbf
  {\bibinfo {volume} {93}},\ \bibinfo {pages} {184301} (\bibinfo {year}
  {2004})}\BibitemShut {NoStop}%
\bibitem [{\citenamefont {Giazotto}\ \emph {et~al.}(2006)\citenamefont
  {Giazotto}, \citenamefont {Heikkil{\"a}}, \citenamefont {Luukanen},
  \citenamefont {Savin},\ and\ \citenamefont
  {Pekola}}]{giazotto2006opportunities}%
  \BibitemOpen
  \bibfield  {author} {\bibinfo {author} {\bibfnamefont {F.}~\bibnamefont
  {Giazotto}}, \bibinfo {author} {\bibfnamefont {T.~T.}\ \bibnamefont
  {Heikkil{\"a}}}, \bibinfo {author} {\bibfnamefont {A.}~\bibnamefont
  {Luukanen}}, \bibinfo {author} {\bibfnamefont {A.~M.}\ \bibnamefont {Savin}},
  \ and\ \bibinfo {author} {\bibfnamefont {J.~P.}\ \bibnamefont {Pekola}},\
  }\href@noop {} {\bibfield  {journal} {\bibinfo  {journal} {Rev. Mod. Phys.}\
  }\textbf {\bibinfo {volume} {78}},\ \bibinfo {pages} {217} (\bibinfo {year}
  {2006})}\BibitemShut {NoStop}%
\bibitem [{\citenamefont {Fornieri}\ and\ \citenamefont
  {Giazotto}(2017)}]{fornieri2017towards}%
  \BibitemOpen
  \bibfield  {author} {\bibinfo {author} {\bibfnamefont {A.}~\bibnamefont
  {Fornieri}}\ and\ \bibinfo {author} {\bibfnamefont {F.}~\bibnamefont
  {Giazotto}},\ }\href@noop {} {\bibfield  {journal} {\bibinfo  {journal} {Nat.
  Nanotechnol.}\ }\textbf {\bibinfo {volume} {12}},\ \bibinfo {pages} {944}
  (\bibinfo {year} {2017})}\BibitemShut {NoStop}%
\bibitem [{\citenamefont {Mart{\'\i}nez-P{\'e}rez}, \citenamefont {Solinas},\
  and\ \citenamefont {Giazotto}(2014)}]{martinez2014coherent}%
  \BibitemOpen
  \bibfield  {author} {\bibinfo {author} {\bibfnamefont {M.}~\bibnamefont
  {Mart{\'\i}nez-P{\'e}rez}}, \bibinfo {author} {\bibfnamefont
  {P.}~\bibnamefont {Solinas}}, \ and\ \bibinfo {author} {\bibfnamefont
  {F.}~\bibnamefont {Giazotto}},\ }\href@noop {} {\bibfield  {journal}
  {\bibinfo  {journal} {J. Low Temp. Phys.}\ }\textbf {\bibinfo {volume}
  {175}},\ \bibinfo {pages} {813} (\bibinfo {year} {2014})}\BibitemShut
  {NoStop}%
\bibitem [{\citenamefont {Fornieri}, \citenamefont {Mart{\'\i}nez-P{\'e}rez},\
  and\ \citenamefont {Giazotto}(2015)}]{fornieri2015electronic}%
  \BibitemOpen
  \bibfield  {author} {\bibinfo {author} {\bibfnamefont {A.}~\bibnamefont
  {Fornieri}}, \bibinfo {author} {\bibfnamefont {M.~J.}\ \bibnamefont
  {Mart{\'\i}nez-P{\'e}rez}}, \ and\ \bibinfo {author} {\bibfnamefont
  {F.}~\bibnamefont {Giazotto}},\ }\href@noop {} {\bibfield  {journal}
  {\bibinfo  {journal} {AIP Adv.}\ }\textbf {\bibinfo {volume} {5}},\ \bibinfo
  {pages} {053301} (\bibinfo {year} {2015})}\BibitemShut {NoStop}%
\bibitem [{\citenamefont {Wu}\ and\ \citenamefont
  {Segal}(2009)}]{wu2009sufficient}%
  \BibitemOpen
  \bibfield  {author} {\bibinfo {author} {\bibfnamefont {L.-A.}\ \bibnamefont
  {Wu}}\ and\ \bibinfo {author} {\bibfnamefont {D.}~\bibnamefont {Segal}},\
  }\href@noop {} {\bibfield  {journal} {\bibinfo  {journal} {Phys. Rev. Lett.}\
  }\textbf {\bibinfo {volume} {102}},\ \bibinfo {pages} {095503} (\bibinfo
  {year} {2009})}\BibitemShut {NoStop}%
\bibitem [{\citenamefont {Segal}(2008)}]{segal2008single}%
  \BibitemOpen
  \bibfield  {author} {\bibinfo {author} {\bibfnamefont {D.}~\bibnamefont
  {Segal}},\ }\href@noop {} {\bibfield  {journal} {\bibinfo  {journal} {Phys.
  Rev. Lett.}\ }\textbf {\bibinfo {volume} {100}},\ \bibinfo {pages} {105901}
  (\bibinfo {year} {2008})}\BibitemShut {NoStop}%
\bibitem [{\citenamefont {Li}, \citenamefont {Wang},\ and\ \citenamefont
  {Casati}(2006)}]{li2006negative}%
  \BibitemOpen
  \bibfield  {author} {\bibinfo {author} {\bibfnamefont {B.}~\bibnamefont
  {Li}}, \bibinfo {author} {\bibfnamefont {L.}~\bibnamefont {Wang}}, \ and\
  \bibinfo {author} {\bibfnamefont {G.}~\bibnamefont {Casati}},\ }\href@noop {}
  {\bibfield  {journal} {\bibinfo  {journal} {Appl. Phys. Lett.}\ }\textbf
  {\bibinfo {volume} {88}},\ \bibinfo {pages} {143501} (\bibinfo {year}
  {2006})}\BibitemShut {NoStop}%
\bibitem [{\citenamefont {Terraneo}, \citenamefont {Peyrard},\ and\
  \citenamefont {Casati}(2002)}]{terraneo2002controlling}%
  \BibitemOpen
  \bibfield  {author} {\bibinfo {author} {\bibfnamefont {M.}~\bibnamefont
  {Terraneo}}, \bibinfo {author} {\bibfnamefont {M.}~\bibnamefont {Peyrard}}, \
  and\ \bibinfo {author} {\bibfnamefont {G.}~\bibnamefont {Casati}},\
  }\href@noop {} {\bibfield  {journal} {\bibinfo  {journal} {Phys. Rev. Lett.}\
  }\textbf {\bibinfo {volume} {88}},\ \bibinfo {pages} {094302} (\bibinfo
  {year} {2002})}\BibitemShut {NoStop}%
\bibitem [{\citenamefont {L{\'o}pez}\ and\ \citenamefont
  {S{\'a}nchez}(2013)}]{lopez2013nonlinear}%
  \BibitemOpen
  \bibfield  {author} {\bibinfo {author} {\bibfnamefont {R.}~\bibnamefont
  {L{\'o}pez}}\ and\ \bibinfo {author} {\bibfnamefont {D.}~\bibnamefont
  {S{\'a}nchez}},\ }\href@noop {} {\bibfield  {journal} {\bibinfo  {journal}
  {Phys. Rev. B}\ }\textbf {\bibinfo {volume} {88}},\ \bibinfo {pages} {045129}
  (\bibinfo {year} {2013})}\BibitemShut {NoStop}%
\bibitem [{\citenamefont {Ren}\ and\ \citenamefont
  {Zhu}(2013)}]{ren2013anomalous}%
  \BibitemOpen
  \bibfield  {author} {\bibinfo {author} {\bibfnamefont {J.}~\bibnamefont
  {Ren}}\ and\ \bibinfo {author} {\bibfnamefont {J.-X.}\ \bibnamefont {Zhu}},\
  }\href@noop {} {\bibfield  {journal} {\bibinfo  {journal} {Phys. Rev. B}\
  }\textbf {\bibinfo {volume} {87}},\ \bibinfo {pages} {165121} (\bibinfo
  {year} {2013})}\BibitemShut {NoStop}%
\bibitem [{\citenamefont {Bours}\ \emph {et~al.}(2019)\citenamefont {Bours},
  \citenamefont {Sothmann}, \citenamefont {Carrega}, \citenamefont {Strambini},
  \citenamefont {Braggio}, \citenamefont {Hankiewicz}, \citenamefont
  {Molenkamp},\ and\ \citenamefont {Giazotto}}]{bours2019phase}%
  \BibitemOpen
  \bibfield  {author} {\bibinfo {author} {\bibfnamefont {L.}~\bibnamefont
  {Bours}}, \bibinfo {author} {\bibfnamefont {B.}~\bibnamefont {Sothmann}},
  \bibinfo {author} {\bibfnamefont {M.}~\bibnamefont {Carrega}}, \bibinfo
  {author} {\bibfnamefont {E.}~\bibnamefont {Strambini}}, \bibinfo {author}
  {\bibfnamefont {A.}~\bibnamefont {Braggio}}, \bibinfo {author} {\bibfnamefont
  {E.~M.}\ \bibnamefont {Hankiewicz}}, \bibinfo {author} {\bibfnamefont
  {L.~W.}\ \bibnamefont {Molenkamp}}, \ and\ \bibinfo {author} {\bibfnamefont
  {F.}~\bibnamefont {Giazotto}},\ }\href@noop {} {\bibfield  {journal}
  {\bibinfo  {journal} {Phys. Rev. Appl.}\ }\textbf {\bibinfo {volume} {11}},\
  \bibinfo {pages} {044073} (\bibinfo {year} {2019})}\BibitemShut {NoStop}%
\bibitem [{\citenamefont {Ruokola}\ and\ \citenamefont
  {Ojanen}(2011)}]{ruokola2011single}%
  \BibitemOpen
  \bibfield  {author} {\bibinfo {author} {\bibfnamefont {T.}~\bibnamefont
  {Ruokola}}\ and\ \bibinfo {author} {\bibfnamefont {T.}~\bibnamefont
  {Ojanen}},\ }\href@noop {} {\bibfield  {journal} {\bibinfo  {journal} {Phys.
  Rev. B}\ }\textbf {\bibinfo {volume} {83}},\ \bibinfo {pages} {241404}
  (\bibinfo {year} {2011})}\BibitemShut {NoStop}%
\bibitem [{\citenamefont {Ruokola}, \citenamefont {Ojanen},\ and\ \citenamefont
  {Jauho}(2009)}]{ruokola2009thermal}%
  \BibitemOpen
  \bibfield  {author} {\bibinfo {author} {\bibfnamefont {T.}~\bibnamefont
  {Ruokola}}, \bibinfo {author} {\bibfnamefont {T.}~\bibnamefont {Ojanen}}, \
  and\ \bibinfo {author} {\bibfnamefont {A.-P.}\ \bibnamefont {Jauho}},\
  }\href@noop {} {\bibfield  {journal} {\bibinfo  {journal} {Phys. Rev. B}\
  }\textbf {\bibinfo {volume} {79}},\ \bibinfo {pages} {144306} (\bibinfo
  {year} {2009})}\BibitemShut {NoStop}%
\bibitem [{\citenamefont {Kuo}\ and\ \citenamefont
  {Chang}(2010)}]{kuo2010thermoelectric}%
  \BibitemOpen
  \bibfield  {author} {\bibinfo {author} {\bibfnamefont {D.~M.-T.}\
  \bibnamefont {Kuo}}\ and\ \bibinfo {author} {\bibfnamefont {Y.-c.}\
  \bibnamefont {Chang}},\ }\href@noop {} {\bibfield  {journal} {\bibinfo
  {journal} {Phys. Rev. B}\ }\textbf {\bibinfo {volume} {81}},\ \bibinfo
  {pages} {205321} (\bibinfo {year} {2010})}\BibitemShut {NoStop}%
\bibitem [{\citenamefont {Fornieri}, \citenamefont {Mart{\'\i}nez-P{\'e}rez},\
  and\ \citenamefont {Giazotto}(2014)}]{fornieri2014normal}%
  \BibitemOpen
  \bibfield  {author} {\bibinfo {author} {\bibfnamefont {A.}~\bibnamefont
  {Fornieri}}, \bibinfo {author} {\bibfnamefont {M.~J.}\ \bibnamefont
  {Mart{\'\i}nez-P{\'e}rez}}, \ and\ \bibinfo {author} {\bibfnamefont
  {F.}~\bibnamefont {Giazotto}},\ }\href@noop {} {\bibfield  {journal}
  {\bibinfo  {journal} {Appl. Phys. Lett.}\ }\textbf {\bibinfo {volume}
  {104}},\ \bibinfo {pages} {183108} (\bibinfo {year} {2014})}\BibitemShut
  {NoStop}%
\bibitem [{\citenamefont {Mart{\'\i}nez-P{\'e}rez}\ and\ \citenamefont
  {Giazotto}(2013)}]{martinez2013efficient}%
  \BibitemOpen
  \bibfield  {author} {\bibinfo {author} {\bibfnamefont {M.}~\bibnamefont
  {Mart{\'\i}nez-P{\'e}rez}}\ and\ \bibinfo {author} {\bibfnamefont
  {F.}~\bibnamefont {Giazotto}},\ }\href@noop {} {\bibfield  {journal}
  {\bibinfo  {journal} {Appl. Phys. Lett.}\ }\textbf {\bibinfo {volume}
  {102}},\ \bibinfo {pages} {182602} (\bibinfo {year} {2013})}\BibitemShut
  {NoStop}%
\bibitem [{\citenamefont {Giazotto}\ and\ \citenamefont
  {Bergeret}(2013)}]{giazotto2013thermal}%
  \BibitemOpen
  \bibfield  {author} {\bibinfo {author} {\bibfnamefont {F.}~\bibnamefont
  {Giazotto}}\ and\ \bibinfo {author} {\bibfnamefont {F.}~\bibnamefont
  {Bergeret}},\ }\href@noop {} {\bibfield  {journal} {\bibinfo  {journal}
  {Appl. Phys. Lett.}\ }\textbf {\bibinfo {volume} {103}},\ \bibinfo {pages}
  {242602} (\bibinfo {year} {2013})}\BibitemShut {NoStop}%
\bibitem [{\citenamefont {Goury}\ and\ \citenamefont
  {S{\'a}nchez}(2019)}]{goury2019reversible}%
  \BibitemOpen
  \bibfield  {author} {\bibinfo {author} {\bibfnamefont {D.}~\bibnamefont
  {Goury}}\ and\ \bibinfo {author} {\bibfnamefont {R.}~\bibnamefont
  {S{\'a}nchez}},\ }\href@noop {} {\bibfield  {journal} {\bibinfo  {journal}
  {Appl. Phys. Lett.}\ }\textbf {\bibinfo {volume} {115}},\ \bibinfo {pages}
  {092601} (\bibinfo {year} {2019})}\BibitemShut {NoStop}%
\bibitem [{\citenamefont {Ben-Abdallah}\ and\ \citenamefont
  {Biehs}(2013)}]{ben2013phase}%
  \BibitemOpen
  \bibfield  {author} {\bibinfo {author} {\bibfnamefont {P.}~\bibnamefont
  {Ben-Abdallah}}\ and\ \bibinfo {author} {\bibfnamefont {S.-A.}\ \bibnamefont
  {Biehs}},\ }\href@noop {} {\bibfield  {journal} {\bibinfo  {journal} {Appl.
  Phys. Lett.}\ }\textbf {\bibinfo {volume} {103}},\ \bibinfo {pages} {191907}
  (\bibinfo {year} {2013})}\BibitemShut {NoStop}%
\bibitem [{\citenamefont {Mart{\'\i}nez-P{\'e}rez}, \citenamefont {Fornieri},\
  and\ \citenamefont {Giazotto}(2015)}]{martinez2015rectification}%
  \BibitemOpen
  \bibfield  {author} {\bibinfo {author} {\bibfnamefont {M.~J.}\ \bibnamefont
  {Mart{\'\i}nez-P{\'e}rez}}, \bibinfo {author} {\bibfnamefont
  {A.}~\bibnamefont {Fornieri}}, \ and\ \bibinfo {author} {\bibfnamefont
  {F.}~\bibnamefont {Giazotto}},\ }\href@noop {} {\bibfield  {journal}
  {\bibinfo  {journal} {Nat. Nanotechnol.}\ }\textbf {\bibinfo {volume} {10}},\
  \bibinfo {pages} {303} (\bibinfo {year} {2015})}\BibitemShut {NoStop}%
\bibitem [{\citenamefont {Senior}\ \emph {et~al.}(2020)\citenamefont {Senior},
  \citenamefont {Gubaydullin}, \citenamefont {Karimi}, \citenamefont
  {Peltonen}, \citenamefont {Ankerhold},\ and\ \citenamefont
  {Pekola}}]{senior2020heat}%
  \BibitemOpen
  \bibfield  {author} {\bibinfo {author} {\bibfnamefont {J.}~\bibnamefont
  {Senior}}, \bibinfo {author} {\bibfnamefont {A.}~\bibnamefont {Gubaydullin}},
  \bibinfo {author} {\bibfnamefont {B.}~\bibnamefont {Karimi}}, \bibinfo
  {author} {\bibfnamefont {J.~T.}\ \bibnamefont {Peltonen}}, \bibinfo {author}
  {\bibfnamefont {J.}~\bibnamefont {Ankerhold}}, \ and\ \bibinfo {author}
  {\bibfnamefont {J.~P.}\ \bibnamefont {Pekola}},\ }\href@noop {} {\bibfield
  {journal} {\bibinfo  {journal} {Comm. Phys.}\ }\textbf {\bibinfo {volume}
  {3}},\ \bibinfo {pages} {1} (\bibinfo {year} {2020})}\BibitemShut {NoStop}%
\bibitem [{\citenamefont {Scheibner}(2008)}]{scheibner2008m}%
  \BibitemOpen
  \bibfield  {author} {\bibinfo {author} {\bibfnamefont {R.}~\bibnamefont
  {Scheibner}},\ }\bibfield  {title} {\enquote {\bibinfo {title} {M. k+ onig,
  d. reuter, ad wieck, c. gould, h. buhmann, lw molenkamp},}\ }\href@noop {}
  {\bibfield  {journal} {\bibinfo  {journal} {New J. Phys}\ }\textbf {\bibinfo
  {volume} {10}},\ \bibinfo {pages} {083016} (\bibinfo {year}
  {2008})}\BibitemShut {NoStop}%
\bibitem [{\citenamefont {Chang}\ \emph {et~al.}(2006)\citenamefont {Chang},
  \citenamefont {Okawa}, \citenamefont {Majumdar},\ and\ \citenamefont
  {Zettl}}]{chang2006solid}%
  \BibitemOpen
  \bibfield  {author} {\bibinfo {author} {\bibfnamefont {C.~W.}\ \bibnamefont
  {Chang}}, \bibinfo {author} {\bibfnamefont {D.}~\bibnamefont {Okawa}},
  \bibinfo {author} {\bibfnamefont {A.}~\bibnamefont {Majumdar}}, \ and\
  \bibinfo {author} {\bibfnamefont {A.}~\bibnamefont {Zettl}},\ }\href@noop {}
  {\bibfield  {journal} {\bibinfo  {journal} {Science}\ }\textbf {\bibinfo
  {volume} {314}},\ \bibinfo {pages} {1121} (\bibinfo {year}
  {2006})}\BibitemShut {NoStop}%
\bibitem [{\citenamefont {Kobayashi}, \citenamefont {Teraoka},\ and\
  \citenamefont {Terasaki}(2009)}]{kobayashi2009oxide}%
  \BibitemOpen
  \bibfield  {author} {\bibinfo {author} {\bibfnamefont {W.}~\bibnamefont
  {Kobayashi}}, \bibinfo {author} {\bibfnamefont {Y.}~\bibnamefont {Teraoka}},
  \ and\ \bibinfo {author} {\bibfnamefont {I.}~\bibnamefont {Terasaki}},\
  }\href@noop {} {\bibfield  {journal} {\bibinfo  {journal} {Appl. Phys.
  Lett.}\ }\textbf {\bibinfo {volume} {95}},\ \bibinfo {pages} {171905}
  (\bibinfo {year} {2009})}\BibitemShut {NoStop}%
\bibitem [{\citenamefont {Tian}\ \emph {et~al.}(2012)\citenamefont {Tian},
  \citenamefont {Xie}, \citenamefont {Yang}, \citenamefont {Ren}, \citenamefont
  {Zhang}, \citenamefont {Wang}, \citenamefont {Zhou}, \citenamefont {Peng},
  \citenamefont {Wang},\ and\ \citenamefont {Liu}}]{tian2012novel}%
  \BibitemOpen
  \bibfield  {author} {\bibinfo {author} {\bibfnamefont {H.}~\bibnamefont
  {Tian}}, \bibinfo {author} {\bibfnamefont {D.}~\bibnamefont {Xie}}, \bibinfo
  {author} {\bibfnamefont {Y.}~\bibnamefont {Yang}}, \bibinfo {author}
  {\bibfnamefont {T.-L.}\ \bibnamefont {Ren}}, \bibinfo {author} {\bibfnamefont
  {G.}~\bibnamefont {Zhang}}, \bibinfo {author} {\bibfnamefont {Y.-F.}\
  \bibnamefont {Wang}}, \bibinfo {author} {\bibfnamefont {C.-J.}\ \bibnamefont
  {Zhou}}, \bibinfo {author} {\bibfnamefont {P.-G.}\ \bibnamefont {Peng}},
  \bibinfo {author} {\bibfnamefont {L.-G.}\ \bibnamefont {Wang}}, \ and\
  \bibinfo {author} {\bibfnamefont {L.-T.}\ \bibnamefont {Liu}},\ }\href@noop
  {} {\bibfield  {journal} {\bibinfo  {journal} {Sci. Rep.}\ }\textbf {\bibinfo
  {volume} {2}},\ \bibinfo {pages} {1} (\bibinfo {year} {2012})}\BibitemShut
  {NoStop}%
\bibitem [{\citenamefont {Bergeret}\ \emph {et~al.}(2018)\citenamefont
  {Bergeret}, \citenamefont {Silaev}, \citenamefont {Virtanen},\ and\
  \citenamefont {Heikkil{\"a}}}]{bergeret2018colloquium}%
  \BibitemOpen
  \bibfield  {author} {\bibinfo {author} {\bibfnamefont {F.~S.}\ \bibnamefont
  {Bergeret}}, \bibinfo {author} {\bibfnamefont {M.}~\bibnamefont {Silaev}},
  \bibinfo {author} {\bibfnamefont {P.}~\bibnamefont {Virtanen}}, \ and\
  \bibinfo {author} {\bibfnamefont {T.~T.}\ \bibnamefont {Heikkil{\"a}}},\
  }\href@noop {} {\bibfield  {journal} {\bibinfo  {journal} {Rev. Mod. Phys.}\
  }\textbf {\bibinfo {volume} {90}},\ \bibinfo {pages} {041001} (\bibinfo
  {year} {2018})}\BibitemShut {NoStop}%
\bibitem [{\citenamefont {Heikkil{\"a}}\ \emph {et~al.}(2019)\citenamefont
  {Heikkil{\"a}}, \citenamefont {Silaev}, \citenamefont {Virtanen},\ and\
  \citenamefont {Bergeret}}]{heikkila2019thermal}%
  \BibitemOpen
  \bibfield  {author} {\bibinfo {author} {\bibfnamefont {T.~T.}\ \bibnamefont
  {Heikkil{\"a}}}, \bibinfo {author} {\bibfnamefont {M.}~\bibnamefont
  {Silaev}}, \bibinfo {author} {\bibfnamefont {P.}~\bibnamefont {Virtanen}}, \
  and\ \bibinfo {author} {\bibfnamefont {F.~S.}\ \bibnamefont {Bergeret}},\
  }\href@noop {} {\bibfield  {journal} {\bibinfo  {journal} {Prog. Surf. Sci.}\
  }\textbf {\bibinfo {volume} {94}},\ \bibinfo {pages} {100540} (\bibinfo
  {year} {2019})}\BibitemShut {NoStop}%
\bibitem [{\citenamefont {Ozaeta}\ \emph {et~al.}(2014)\citenamefont {Ozaeta},
  \citenamefont {Virtanen}, \citenamefont {Bergeret},\ and\ \citenamefont
  {Heikkil{\"a}}}]{ozaeta2014predicted}%
  \BibitemOpen
  \bibfield  {author} {\bibinfo {author} {\bibfnamefont {A.}~\bibnamefont
  {Ozaeta}}, \bibinfo {author} {\bibfnamefont {P.}~\bibnamefont {Virtanen}},
  \bibinfo {author} {\bibfnamefont {F.}~\bibnamefont {Bergeret}}, \ and\
  \bibinfo {author} {\bibfnamefont {T.}~\bibnamefont {Heikkil{\"a}}},\
  }\href@noop {} {\bibfield  {journal} {\bibinfo  {journal} {Phys. Rev. Lett.}\
  }\textbf {\bibinfo {volume} {112}},\ \bibinfo {pages} {057001} (\bibinfo
  {year} {2014})}\BibitemShut {NoStop}%
\bibitem [{\citenamefont {Machon}, \citenamefont {Eschrig},\ and\ \citenamefont
  {Belzig}(2013)}]{machon2013nonlocal}%
  \BibitemOpen
  \bibfield  {author} {\bibinfo {author} {\bibfnamefont {P.}~\bibnamefont
  {Machon}}, \bibinfo {author} {\bibfnamefont {M.}~\bibnamefont {Eschrig}}, \
  and\ \bibinfo {author} {\bibfnamefont {W.}~\bibnamefont {Belzig}},\
  }\href@noop {} {\bibfield  {journal} {\bibinfo  {journal} {Phys. Rev. Lett.}\
  }\textbf {\bibinfo {volume} {110}},\ \bibinfo {pages} {047002} (\bibinfo
  {year} {2013})}\BibitemShut {NoStop}%
\bibitem [{\citenamefont {Machon}, \citenamefont {Eschrig},\ and\ \citenamefont
  {Belzig}(2014)}]{machon2014giant}%
  \BibitemOpen
  \bibfield  {author} {\bibinfo {author} {\bibfnamefont {P.}~\bibnamefont
  {Machon}}, \bibinfo {author} {\bibfnamefont {M.}~\bibnamefont {Eschrig}}, \
  and\ \bibinfo {author} {\bibfnamefont {W.}~\bibnamefont {Belzig}},\
  }\href@noop {} {\bibfield  {journal} {\bibinfo  {journal} {New J. Phys.}\
  }\textbf {\bibinfo {volume} {16}},\ \bibinfo {pages} {073002} (\bibinfo
  {year} {2014})}\BibitemShut {NoStop}%
\bibitem [{\citenamefont {Kolenda}, \citenamefont {Wolf},\ and\ \citenamefont
  {Beckmann}(2016)}]{kolenda2016observation}%
  \BibitemOpen
  \bibfield  {author} {\bibinfo {author} {\bibfnamefont {S.}~\bibnamefont
  {Kolenda}}, \bibinfo {author} {\bibfnamefont {M.~J.}\ \bibnamefont {Wolf}}, \
  and\ \bibinfo {author} {\bibfnamefont {D.}~\bibnamefont {Beckmann}},\
  }\href@noop {} {\bibfield  {journal} {\bibinfo  {journal} {Phys. Rev. Lett.}\
  }\textbf {\bibinfo {volume} {116}},\ \bibinfo {pages} {097001} (\bibinfo
  {year} {2016})}\BibitemShut {NoStop}%
\bibitem [{\citenamefont {Linder}\ and\ \citenamefont
  {Robinson}(2015)}]{linder2015superconducting}%
  \BibitemOpen
  \bibfield  {author} {\bibinfo {author} {\bibfnamefont {J.}~\bibnamefont
  {Linder}}\ and\ \bibinfo {author} {\bibfnamefont {J.~W.}\ \bibnamefont
  {Robinson}},\ }\href@noop {} {\bibfield  {journal} {\bibinfo  {journal} {Nat.
  Phys.}\ }\textbf {\bibinfo {volume} {11}},\ \bibinfo {pages} {307} (\bibinfo
  {year} {2015})}\BibitemShut {NoStop}%
\bibitem [{\citenamefont {Tokuyasu}, \citenamefont {Sauls},\ and\ \citenamefont
  {Rainer}(1988)}]{tokuyasu1988proximity}%
  \BibitemOpen
  \bibfield  {author} {\bibinfo {author} {\bibfnamefont {T.}~\bibnamefont
  {Tokuyasu}}, \bibinfo {author} {\bibfnamefont {J.~A.}\ \bibnamefont {Sauls}},
  \ and\ \bibinfo {author} {\bibfnamefont {D.}~\bibnamefont {Rainer}},\
  }\href@noop {} {\bibfield  {journal} {\bibinfo  {journal} {Phys. Rev. B}\
  }\textbf {\bibinfo {volume} {38}},\ \bibinfo {pages} {8823} (\bibinfo {year}
  {1988})}\BibitemShut {NoStop}%
\bibitem [{\citenamefont {Dynes}\ \emph {et~al.}(1984)\citenamefont {Dynes},
  \citenamefont {Garno}, \citenamefont {Hertel},\ and\ \citenamefont
  {Orlando}}]{dynes1984tunneling}%
  \BibitemOpen
  \bibfield  {author} {\bibinfo {author} {\bibfnamefont {R.}~\bibnamefont
  {Dynes}}, \bibinfo {author} {\bibfnamefont {J.}~\bibnamefont {Garno}},
  \bibinfo {author} {\bibfnamefont {G.}~\bibnamefont {Hertel}}, \ and\ \bibinfo
  {author} {\bibfnamefont {T.}~\bibnamefont {Orlando}},\ }\href@noop {}
  {\bibfield  {journal} {\bibinfo  {journal} {Phys. Rev. Lett.}\ }\textbf
  {\bibinfo {volume} {53}},\ \bibinfo {pages} {2437} (\bibinfo {year}
  {1984})}\BibitemShut {NoStop}%
\bibitem [{\citenamefont {Moodera}, \citenamefont {Santos},\ and\ \citenamefont
  {Nagahama}(2007)}]{moodera2007phenomena}%
  \BibitemOpen
  \bibfield  {author} {\bibinfo {author} {\bibfnamefont {J.~S.}\ \bibnamefont
  {Moodera}}, \bibinfo {author} {\bibfnamefont {T.~S.}\ \bibnamefont {Santos}},
  \ and\ \bibinfo {author} {\bibfnamefont {T.}~\bibnamefont {Nagahama}},\
  }\href@noop {} {\bibfield  {journal} {\bibinfo  {journal} {J. Phys.: Condens.
  Matter}\ }\textbf {\bibinfo {volume} {19}},\ \bibinfo {pages} {165202}
  (\bibinfo {year} {2007})}\BibitemShut {NoStop}%
\bibitem [{\citenamefont {Giazotto}\ and\ \citenamefont
  {Mart{\'\i}nez-P{\'e}rez}(2012)}]{giazotto2012josephson}%
  \BibitemOpen
  \bibfield  {author} {\bibinfo {author} {\bibfnamefont {F.}~\bibnamefont
  {Giazotto}}\ and\ \bibinfo {author} {\bibfnamefont {M.~J.}\ \bibnamefont
  {Mart{\'\i}nez-P{\'e}rez}},\ }\href@noop {} {\bibfield  {journal} {\bibinfo
  {journal} {Nature}\ }\textbf {\bibinfo {volume} {492}},\ \bibinfo {pages}
  {401} (\bibinfo {year} {2012})}\BibitemShut {NoStop}%
\bibitem [{\citenamefont {Meservey}\ and\ \citenamefont
  {Tedrow}(1994)}]{meservey1994spin}%
  \BibitemOpen
  \bibfield  {author} {\bibinfo {author} {\bibfnamefont {R.}~\bibnamefont
  {Meservey}}\ and\ \bibinfo {author} {\bibfnamefont {P.}~\bibnamefont
  {Tedrow}},\ }\href@noop {} {\bibfield  {journal} {\bibinfo  {journal} {Phys.
  Rep.}\ }\textbf {\bibinfo {volume} {238}},\ \bibinfo {pages} {173} (\bibinfo
  {year} {1994})}\BibitemShut {NoStop}%
\bibitem [{\citenamefont {Hao}, \citenamefont {Moodera},\ and\ \citenamefont
  {Meservey}(1990)}]{hao1990spin}%
  \BibitemOpen
  \bibfield  {author} {\bibinfo {author} {\bibfnamefont {X.}~\bibnamefont
  {Hao}}, \bibinfo {author} {\bibfnamefont {J.}~\bibnamefont {Moodera}}, \ and\
  \bibinfo {author} {\bibfnamefont {R.}~\bibnamefont {Meservey}},\ }\href@noop
  {} {\bibfield  {journal} {\bibinfo  {journal} {Phys. Rev. B}\ }\textbf
  {\bibinfo {volume} {42}},\ \bibinfo {pages} {8235} (\bibinfo {year}
  {1990})}\BibitemShut {NoStop}%
\bibitem [{\citenamefont {Li}, \citenamefont {Miao},\ and\ \citenamefont
  {Moodera}(2013)}]{li2013observation}%
  \BibitemOpen
  \bibfield  {author} {\bibinfo {author} {\bibfnamefont {B.}~\bibnamefont
  {Li}}, \bibinfo {author} {\bibfnamefont {G.-X.}\ \bibnamefont {Miao}}, \ and\
  \bibinfo {author} {\bibfnamefont {J.~S.}\ \bibnamefont {Moodera}},\
  }\href@noop {} {\bibfield  {journal} {\bibinfo  {journal} {Phys. Rev. B}\
  }\textbf {\bibinfo {volume} {88}},\ \bibinfo {pages} {161105} (\bibinfo
  {year} {2013})}\BibitemShut {NoStop}%
\bibitem [{\citenamefont {Li}\ \emph {et~al.}(2013)\citenamefont {Li},
  \citenamefont {Roschewsky}, \citenamefont {Assaf}, \citenamefont {Eich},
  \citenamefont {Epstein-Martin}, \citenamefont {Heiman}, \citenamefont
  {M{\"u}nzenberg},\ and\ \citenamefont {Moodera}}]{li2013superconducting}%
  \BibitemOpen
  \bibfield  {author} {\bibinfo {author} {\bibfnamefont {B.}~\bibnamefont
  {Li}}, \bibinfo {author} {\bibfnamefont {N.}~\bibnamefont {Roschewsky}},
  \bibinfo {author} {\bibfnamefont {B.~A.}\ \bibnamefont {Assaf}}, \bibinfo
  {author} {\bibfnamefont {M.}~\bibnamefont {Eich}}, \bibinfo {author}
  {\bibfnamefont {M.}~\bibnamefont {Epstein-Martin}}, \bibinfo {author}
  {\bibfnamefont {D.}~\bibnamefont {Heiman}}, \bibinfo {author} {\bibfnamefont
  {M.}~\bibnamefont {M{\"u}nzenberg}}, \ and\ \bibinfo {author} {\bibfnamefont
  {J.~S.}\ \bibnamefont {Moodera}},\ }\href@noop {} {\bibfield  {journal}
  {\bibinfo  {journal} {Phys. Rev. Lett.}\ }\textbf {\bibinfo {volume} {110}},\
  \bibinfo {pages} {097001} (\bibinfo {year} {2013})}\BibitemShut {NoStop}%
\bibitem [{\citenamefont {Xiong}\ \emph {et~al.}(2011)\citenamefont {Xiong},
  \citenamefont {Stadler}, \citenamefont {Adams},\ and\ \citenamefont
  {Catelani}}]{xiong2011spin}%
  \BibitemOpen
  \bibfield  {author} {\bibinfo {author} {\bibfnamefont {Y.}~\bibnamefont
  {Xiong}}, \bibinfo {author} {\bibfnamefont {S.}~\bibnamefont {Stadler}},
  \bibinfo {author} {\bibfnamefont {P.}~\bibnamefont {Adams}}, \ and\ \bibinfo
  {author} {\bibfnamefont {G.}~\bibnamefont {Catelani}},\ }\href@noop {}
  {\bibfield  {journal} {\bibinfo  {journal} {Phys. Rev. Lett.}\ }\textbf
  {\bibinfo {volume} {106}},\ \bibinfo {pages} {247001} (\bibinfo {year}
  {2011})}\BibitemShut {NoStop}%
\bibitem [{\citenamefont {Liu}, \citenamefont {Prestigiacomo},\ and\
  \citenamefont {Adams}(2013)}]{liu2013electrostatic}%
  \BibitemOpen
  \bibfield  {author} {\bibinfo {author} {\bibfnamefont {T.}~\bibnamefont
  {Liu}}, \bibinfo {author} {\bibfnamefont {J.}~\bibnamefont {Prestigiacomo}},
  \ and\ \bibinfo {author} {\bibfnamefont {P.}~\bibnamefont {Adams}},\
  }\href@noop {} {\bibfield  {journal} {\bibinfo  {journal} {Phys. Rev. Lett.}\
  }\textbf {\bibinfo {volume} {111}},\ \bibinfo {pages} {027207} (\bibinfo
  {year} {2013})}\BibitemShut {NoStop}%
\bibitem [{\citenamefont {Wolf}\ \emph {et~al.}(2014)\citenamefont {Wolf},
  \citenamefont {S{\"u}rgers}, \citenamefont {Fischer},\ and\ \citenamefont
  {Beckmann}}]{wolf2014spin}%
  \BibitemOpen
  \bibfield  {author} {\bibinfo {author} {\bibfnamefont {M.}~\bibnamefont
  {Wolf}}, \bibinfo {author} {\bibfnamefont {C.}~\bibnamefont {S{\"u}rgers}},
  \bibinfo {author} {\bibfnamefont {G.}~\bibnamefont {Fischer}}, \ and\
  \bibinfo {author} {\bibfnamefont {D.}~\bibnamefont {Beckmann}},\ }\href@noop
  {} {\bibfield  {journal} {\bibinfo  {journal} {Phys. Rev. B}\ }\textbf
  {\bibinfo {volume} {90}},\ \bibinfo {pages} {144509} (\bibinfo {year}
  {2014})}\BibitemShut {NoStop}%
\bibitem [{\citenamefont {Strambini}\ \emph {et~al.}(2017)\citenamefont
  {Strambini}, \citenamefont {Golovach}, \citenamefont {De~Simoni},
  \citenamefont {Moodera}, \citenamefont {Bergeret},\ and\ \citenamefont
  {Giazotto}}]{strambini2017revealing}%
  \BibitemOpen
  \bibfield  {author} {\bibinfo {author} {\bibfnamefont {E.}~\bibnamefont
  {Strambini}}, \bibinfo {author} {\bibfnamefont {V.}~\bibnamefont {Golovach}},
  \bibinfo {author} {\bibfnamefont {G.}~\bibnamefont {De~Simoni}}, \bibinfo
  {author} {\bibfnamefont {J.}~\bibnamefont {Moodera}}, \bibinfo {author}
  {\bibfnamefont {F.}~\bibnamefont {Bergeret}}, \ and\ \bibinfo {author}
  {\bibfnamefont {F.}~\bibnamefont {Giazotto}},\ }\href@noop {} {\bibfield
  {journal} {\bibinfo  {journal} {Phys. Rev. Mater.}\ }\textbf {\bibinfo
  {volume} {1}},\ \bibinfo {pages} {054402} (\bibinfo {year}
  {2017})}\BibitemShut {NoStop}%
\bibitem [{\citenamefont {De~Simoni}\ \emph {et~al.}(2018)\citenamefont
  {De~Simoni}, \citenamefont {Strambini}, \citenamefont {Moodera},
  \citenamefont {Bergeret},\ and\ \citenamefont {Giazotto}}]{de2018toward}%
  \BibitemOpen
  \bibfield  {author} {\bibinfo {author} {\bibfnamefont {G.}~\bibnamefont
  {De~Simoni}}, \bibinfo {author} {\bibfnamefont {E.}~\bibnamefont
  {Strambini}}, \bibinfo {author} {\bibfnamefont {J.~S.}\ \bibnamefont
  {Moodera}}, \bibinfo {author} {\bibfnamefont {F.~S.}\ \bibnamefont
  {Bergeret}}, \ and\ \bibinfo {author} {\bibfnamefont {F.}~\bibnamefont
  {Giazotto}},\ }\href@noop {} {\bibfield  {journal} {\bibinfo  {journal} {Nano
  Lett.}\ }\textbf {\bibinfo {volume} {18}},\ \bibinfo {pages} {6369} (\bibinfo
  {year} {2018})}\BibitemShut {NoStop}%
\bibitem [{\citenamefont {Rouco}\ \emph {et~al.}(2019)\citenamefont {Rouco},
  \citenamefont {Chakraborty}, \citenamefont {Aikebaier}, \citenamefont
  {Golovach}, \citenamefont {Strambini}, \citenamefont {Moodera}, \citenamefont
  {Giazotto}, \citenamefont {Heikkil{\"a}},\ and\ \citenamefont
  {Bergeret}}]{rouco2019charge}%
  \BibitemOpen
  \bibfield  {author} {\bibinfo {author} {\bibfnamefont {M.}~\bibnamefont
  {Rouco}}, \bibinfo {author} {\bibfnamefont {S.}~\bibnamefont {Chakraborty}},
  \bibinfo {author} {\bibfnamefont {F.}~\bibnamefont {Aikebaier}}, \bibinfo
  {author} {\bibfnamefont {V.~N.}\ \bibnamefont {Golovach}}, \bibinfo {author}
  {\bibfnamefont {E.}~\bibnamefont {Strambini}}, \bibinfo {author}
  {\bibfnamefont {J.~S.}\ \bibnamefont {Moodera}}, \bibinfo {author}
  {\bibfnamefont {F.}~\bibnamefont {Giazotto}}, \bibinfo {author}
  {\bibfnamefont {T.~T.}\ \bibnamefont {Heikkil{\"a}}}, \ and\ \bibinfo
  {author} {\bibfnamefont {F.~S.}\ \bibnamefont {Bergeret}},\ }\href@noop {}
  {\bibfield  {journal} {\bibinfo  {journal} {Phys. Rev. B}\ }\textbf {\bibinfo
  {volume} {100}},\ \bibinfo {pages} {184501} (\bibinfo {year}
  {2019})}\BibitemShut {NoStop}%
\bibitem [{\citenamefont {Giazotto}\ \emph {et~al.}(2015)\citenamefont
  {Giazotto}, \citenamefont {Solinas}, \citenamefont {Braggio},\ and\
  \citenamefont {Bergeret}}]{giazotto2015ferromagnetic}%
  \BibitemOpen
  \bibfield  {author} {\bibinfo {author} {\bibfnamefont {F.}~\bibnamefont
  {Giazotto}}, \bibinfo {author} {\bibfnamefont {P.}~\bibnamefont {Solinas}},
  \bibinfo {author} {\bibfnamefont {A.}~\bibnamefont {Braggio}}, \ and\
  \bibinfo {author} {\bibfnamefont {F.~S.}\ \bibnamefont {Bergeret}},\
  }\href@noop {} {\bibfield  {journal} {\bibinfo  {journal} {Phys. Rev. Appl.}\
  }\textbf {\bibinfo {volume} {4}},\ \bibinfo {pages} {044016} (\bibinfo {year}
  {2015})}\BibitemShut {NoStop}%
\bibitem [{\citenamefont {Moodera}, \citenamefont {Miao},\ and\ \citenamefont
  {Santos}(2010)}]{moodera2010frontiers}%
  \BibitemOpen
  \bibfield  {author} {\bibinfo {author} {\bibfnamefont {J.~S.}\ \bibnamefont
  {Moodera}}, \bibinfo {author} {\bibfnamefont {G.-X.}\ \bibnamefont {Miao}}, \
  and\ \bibinfo {author} {\bibfnamefont {T.~S.}\ \bibnamefont {Santos}},\
  }\href@noop {} {\bibfield  {journal} {\bibinfo  {journal} {Phys. Today}\
  }\textbf {\bibinfo {volume} {63}},\ \bibinfo {pages} {46} (\bibinfo {year}
  {2010})}\BibitemShut {NoStop}%
\bibitem [{\citenamefont {Moodera}\ \emph {et~al.}(1988)\citenamefont
  {Moodera}, \citenamefont {Hao}, \citenamefont {Gibson},\ and\ \citenamefont
  {Meservey}}]{moodera1988electron}%
  \BibitemOpen
  \bibfield  {author} {\bibinfo {author} {\bibfnamefont {J.}~\bibnamefont
  {Moodera}}, \bibinfo {author} {\bibfnamefont {X.}~\bibnamefont {Hao}},
  \bibinfo {author} {\bibfnamefont {G.}~\bibnamefont {Gibson}}, \ and\ \bibinfo
  {author} {\bibfnamefont {R.}~\bibnamefont {Meservey}},\ }\href@noop {}
  {\bibfield  {journal} {\bibinfo  {journal} {Phys. Rev. Lett.}\ }\textbf
  {\bibinfo {volume} {61}},\ \bibinfo {pages} {637} (\bibinfo {year}
  {1988})}\BibitemShut {NoStop}%
\bibitem [{\citenamefont {Santos}\ and\ \citenamefont
  {Moodera}(2004)}]{santos2004observation}%
  \BibitemOpen
  \bibfield  {author} {\bibinfo {author} {\bibfnamefont {T.~S.}\ \bibnamefont
  {Santos}}\ and\ \bibinfo {author} {\bibfnamefont {J.~S.}\ \bibnamefont
  {Moodera}},\ }\href@noop {} {\bibfield  {journal} {\bibinfo  {journal} {Phys.
  Rev. B}\ }\textbf {\bibinfo {volume} {69}},\ \bibinfo {pages} {241203}
  (\bibinfo {year} {2004})}\BibitemShut {NoStop}%
\bibitem [{\citenamefont {Moodera}, \citenamefont {Meservey},\ and\
  \citenamefont {Hao}(1993)}]{moodera1993variation}%
  \BibitemOpen
  \bibfield  {author} {\bibinfo {author} {\bibfnamefont {J.}~\bibnamefont
  {Moodera}}, \bibinfo {author} {\bibfnamefont {R.}~\bibnamefont {Meservey}}, \
  and\ \bibinfo {author} {\bibfnamefont {X.}~\bibnamefont {Hao}},\ }\href@noop
  {} {\bibfield  {journal} {\bibinfo  {journal} {Phys. Rev. Lett.}\ }\textbf
  {\bibinfo {volume} {70}},\ \bibinfo {pages} {853} (\bibinfo {year}
  {1993})}\BibitemShut {NoStop}%
\bibitem [{\citenamefont {Santos}\ \emph {et~al.}(2008)\citenamefont {Santos},
  \citenamefont {Moodera}, \citenamefont {Raman}, \citenamefont {Negusse},
  \citenamefont {Holroyd}, \citenamefont {Dvorak}, \citenamefont {Liberati},
  \citenamefont {Idzerda},\ and\ \citenamefont
  {Arenholz}}]{santos2008determining}%
  \BibitemOpen
  \bibfield  {author} {\bibinfo {author} {\bibfnamefont {T.}~\bibnamefont
  {Santos}}, \bibinfo {author} {\bibfnamefont {J.}~\bibnamefont {Moodera}},
  \bibinfo {author} {\bibfnamefont {K.}~\bibnamefont {Raman}}, \bibinfo
  {author} {\bibfnamefont {E.}~\bibnamefont {Negusse}}, \bibinfo {author}
  {\bibfnamefont {J.}~\bibnamefont {Holroyd}}, \bibinfo {author} {\bibfnamefont
  {J.}~\bibnamefont {Dvorak}}, \bibinfo {author} {\bibfnamefont
  {M.}~\bibnamefont {Liberati}}, \bibinfo {author} {\bibfnamefont
  {Y.}~\bibnamefont {Idzerda}}, \ and\ \bibinfo {author} {\bibfnamefont
  {E.}~\bibnamefont {Arenholz}},\ }\href@noop {} {\bibfield  {journal}
  {\bibinfo  {journal} {Phys. Rev. Lett.}\ }\textbf {\bibinfo {volume} {101}},\
  \bibinfo {pages} {147201} (\bibinfo {year} {2008})}\BibitemShut {NoStop}%
\bibitem [{\citenamefont {Miao}\ and\ \citenamefont
  {Moodera}(2009)}]{miao2009controlling}%
  \BibitemOpen
  \bibfield  {author} {\bibinfo {author} {\bibfnamefont {G.-X.}\ \bibnamefont
  {Miao}}\ and\ \bibinfo {author} {\bibfnamefont {J.~S.}\ \bibnamefont
  {Moodera}},\ }\href@noop {} {\bibfield  {journal} {\bibinfo  {journal} {Appl.
  Phys. Lett.}\ }\textbf {\bibinfo {volume} {94}},\ \bibinfo {pages} {182504}
  (\bibinfo {year} {2009})}\BibitemShut {NoStop}%
\bibitem [{\citenamefont {Senapati}, \citenamefont {Blamire},\ and\
  \citenamefont {Barber}(2011)}]{senapati2011spin}%
  \BibitemOpen
  \bibfield  {author} {\bibinfo {author} {\bibfnamefont {K.}~\bibnamefont
  {Senapati}}, \bibinfo {author} {\bibfnamefont {M.~G.}\ \bibnamefont
  {Blamire}}, \ and\ \bibinfo {author} {\bibfnamefont {Z.~H.}\ \bibnamefont
  {Barber}},\ }\href@noop {} {\bibfield  {journal} {\bibinfo  {journal} {Nat.
  Mater.}\ }\textbf {\bibinfo {volume} {10}},\ \bibinfo {pages} {849} (\bibinfo
  {year} {2011})}\BibitemShut {NoStop}%
\bibitem [{\citenamefont {Pal}\ \emph {et~al.}(2013)\citenamefont {Pal},
  \citenamefont {Senapati}, \citenamefont {Barber},\ and\ \citenamefont
  {Blamire}}]{pal2013electric}%
  \BibitemOpen
  \bibfield  {author} {\bibinfo {author} {\bibfnamefont {A.}~\bibnamefont
  {Pal}}, \bibinfo {author} {\bibfnamefont {K.}~\bibnamefont {Senapati}},
  \bibinfo {author} {\bibfnamefont {Z.}~\bibnamefont {Barber}}, \ and\ \bibinfo
  {author} {\bibfnamefont {M.}~\bibnamefont {Blamire}},\ }\href@noop {}
  {\bibfield  {journal} {\bibinfo  {journal} {Adv. Mater.}\ }\textbf {\bibinfo
  {volume} {25}},\ \bibinfo {pages} {5581} (\bibinfo {year}
  {2013})}\BibitemShut {NoStop}%
\bibitem [{\citenamefont {Pal}\ \emph {et~al.}(2014)\citenamefont {Pal},
  \citenamefont {Barber}, \citenamefont {Robinson},\ and\ \citenamefont
  {Blamire}}]{pal2014pure}%
  \BibitemOpen
  \bibfield  {author} {\bibinfo {author} {\bibfnamefont {A.}~\bibnamefont
  {Pal}}, \bibinfo {author} {\bibfnamefont {Z.}~\bibnamefont {Barber}},
  \bibinfo {author} {\bibfnamefont {J.}~\bibnamefont {Robinson}}, \ and\
  \bibinfo {author} {\bibfnamefont {M.}~\bibnamefont {Blamire}},\ }\href@noop
  {} {\bibfield  {journal} {\bibinfo  {journal} {Nat. Commun.}\ }\textbf
  {\bibinfo {volume} {5}},\ \bibinfo {pages} {1} (\bibinfo {year}
  {2014})}\BibitemShut {NoStop}%
\end{thebibliography}
%

\end{document}